\documentclass[12pt,authoryear]{elsarticle}

\usepackage{amssymb}

\usepackage{array}
\usepackage{pdflscape}
\usepackage{longtable}

\usepackage[acronym]{glossaries}

\newcommand{\rqoneOccur}{RQ1: What experiences lead to psychological overwhelm within software development activities?}
\newcommand{\rqoneOccurShort}{What led to psychological overwhelm?}
\newcommand{\rqtwoExperience}{RQ2: How do software developers experience overwhelm in their daily work?}
\newcommand{\rqtwoExperienceShort}{How was overwhelm experienced?}
\newcommand{\rqthreeProductivity}{RQ3: How do software developers experience overwhelm in relation to their productivity?}
\newcommand{\rqthreeProductivityShort}{How was productivity affected?}
\newcommand{\rqfourStress}{RQ4: What role does stress play in developer's experience of overwhelm?}
\newcommand{\rqfourStressShort}{What role does stress play?}

\newacronym{ipa}{IPA}{Interpretive Phenomenological Analysis}

\journal{\ }

\begin{document}

\begin{frontmatter}

\title{Overwhelmed software developers: An Interpretative Phenomenological
Analysis}

\author[inst1,inst2]{Lisa-Marie Michels}
\author[inst1]{Aleksandra Petkova}
\author[inst1]{Marcel Richter}
\author[inst1]{Andreas Farley}
\author[inst1]{Daniel Graziotin}
\author[inst1]{Stefan Wagner}

\affiliation[inst1]{organization={Institute of Software Engineering, University of Stuttgart},%
addressline={Universitätsstraße 38}, 
city={Stuttgart},
postcode={70569}, 
state={BW},
country={Germany}}

\affiliation[inst2]{contact={Contact: lmichels.research@gmail.com}}

\begin{abstract}
\glsresetall
In this paper, we report on an~\gls{ipa} study on experiencing overwhelm in a software development context. The \textit{objectives} of our study are, hence, to understand the experiences developers have when being overwhelmed, how this impacts their productivity and which role stress plays in the process. To this end, we interviewed two software developers who have experienced overwhelm recently. Throughout a qualitative analysis of the shared experiences, we uncover seven categories of overwhelm (communication, disturbance, organizational, variety, technical, temporal, and  positive overwhelm). While the first six themes all are related to negative outcomes, including low productivity and stress, the participants reported that overwhelm can sometimes be experienced to be positive and pleasant, and it can increase their mental focus, self ambition, and productivity. Stress was the most mentioned feeling experienced when overwhelmed. Our findings, for the most, are along the same direction of similar studies from other disciplines and with other participants. However, there may be unique attributes to software developers that mitigate the negative experiences of overwhelm.  
\end{abstract}

\glsresetall


\end{frontmatter}

\newpage
\section{Introduction}\label{sec:Introduction}
\glsresetall
One of the defining traits that may separate human beings from machines is the ability to feel emotions \citep{picard1999affective}. This trait can benefit us greatly, with emotions having been found to increase efficiency and promote more creative problem-solving capabilities~\citep{Hirt1996ProcessingGT,TheRoleofAffectiveExperience}. The frequent experiencing of positive emotions has been directly linked to productivity in the workplace~\citep{oswald2015happiness}. Positive emotions bolster creative thinking and encourage the exchange of ideas~\citep{oswald2015happiness,TheEmergingRoleOfeEmotionsInWorkLife} and have been correlated to increased productivity in the software engineering domain~\citep{AreHappyDevelopersMoreProductive}. 

Emotions, however, are not always positive, and negative emotions have mostly been found to be associated with detrimental aspects of software engineering such as low code quality, work withdrawal, low cognitive performance~\citep{graziotin2018What} and, in general, to lower productivity \citep{EmotionsInTheSoftwareDevelopmentProcess,graziotin2015How}. Negative emotions are not necessarily detrimental, e.g., anger stemming from conflicts among software developers may be the spark needed to find the solution to a problem\citep{AgileSoftwareTeamsCanUseConflict}. The consensus in the software engineering discipline, so far, is that we should strive to limit negative experiences and the resulting unhappiness~\citep{HappinessAndTheProductivityOfSoftwareEngineers}.

One phenomenon induced by negative emotions is psychological overwhelm.\footnote{In this paper, we use the term \textit{overwhelm} as a noun to mean ``The act of overwhelming, or fact of being overwhelmed.''\citep{oed:overwhelm} in line with popular books on the subject, e.g., \citep{van2018age,lipskyAgeOverwhelmedStrategies2018}.} As put by~\cite[p. 52]{van2018age}, ``The state of overwhelm has many shades. It is a continuum. But many of us are, in fact, experiencing some degree of overwhelm as a natural response to all that we encounter---whether we experience that as an occasional flutter of doubt, more frequent flooding of emotion, or gripping despair that we carry with us throughout our days.''. Overwhelm is, indeed, a phenomenon all of us can relate to, it is a universal living experience~\citep{LefdahlDavis2019}; yet, despite being experienced and lived by all, overwhelm has only been studied by few scholars \citep{FeelingOverwhelmedAParsesciencingInquiry}. 

The occurrence of psychological overwhelm is not instantaneous, but rather the result of moment to moment decisions stemming from human emotions,  moods, desires, and intentions \citep{FeelingOverwhelmedAParsesciencingInquiry}. This aspect of overwhelm makes it difficult to find a singular, all encompassing definition of overwhelm,\footnote{See Section~\ref{ssec:overwhelm} for a working definition of overwhelm.}, which has led scholars to form emerging themes from different discipline-related definitions~\citep{doi:10.1177/0894318418807931}. While the impact of overwhelm on different disciplines such as social work, theology, philosophy, psychology, business, nursing, education, and sociology has been explored, overwhelm in connection to software engineering is an unexplored avenue of research. Due to the aforementioned connection of emotions with productivity, psychological overwhelm of software developers may impact productivity, a connection which we explore in this paper.

The \textit{objectives} of our study are, hence, to understand the experiences developers have when being overwhelmed, how this impacts their productivity and which role stress plays in the process.

Our quest to investigate the experiencing of phenomena in software engineering has been met in a recent call for research by~\cite{lenberg2023Qualitative} in the behavioral software engineering sub-discipline. 

\citet{lenberg2023Qualitative} explains that interpretative phenomenological analysis (IPA), a methodology commonly used in psychology, focuses on individuals' personal interpretation of their social experiences. 

This approach, which was established in the mid-90s, is now applicable across numerous strands of psychology.
IPA is grounded in the central principles of phenomenology, hermeneutics, and ideography. Phenomenological studies accentuate individuals' perceived experiences. It employs the concept of 'lifeworld'—the subjective experience of self, objects, and relationships—which phenomenologists strive to grasp by setting aside preconceptions through a process called \textit{epoché} or phenomenological reduction.

Hermeneutics interprets the meaning uncovered by phenomenology, while ideography focuses on the fine-grained analysis of individual cases. IPA studies typically start with broad, exploratory research questions, often carried out on small, purposively selected samples.

Lenberg et al.\ proposes that IPA's detailed exploration of individual experiences makes it an excellent tool for understanding different aspects of software engineering. It uncovers developers' perspectives and thought processes, aiding in comprehending challenges and finding solutions that incorporate human elements. IPA can also balance the group-centric focus of modern agile software teams by examining the unique experiences of individual software engineers~\citep{lenberg2023Qualitative}.

The authors suggest potential IPA applications in software engineering research, such as investigating team dynamics in agile environments, understanding the experiences of adopting new technologies, and examining the effects of stress and mental health issues on software developers. We invite readers to their paper for learning more about IPA.

Hence, and based on the guidelines by~\citet{lenberg2023Qualitative}, we make use of a qualitative research method which, to the best of our knowledge, has not yet been used in software engineering research. We identify~\gls{ipa}~\citep{InterpretativePhenomenologicalAnalysis} to be a suitable research method to fulfill our aims. As overwhelm is often described through themes emerging from interviews reporting on experiences, the individual life experiences and detailed personal accounts provided through~\gls{ipa} are suitable to form our own emerging themes and correlate them to personal accounts of productivity~\citep{ApracticalguidetousingInterpretativePhenomenologicalAnalysis}.  

The occurrence of psychological overwhelm is the culmination of a chain of experiences that leads a human to becoming overwhelmed~\citep{doi:10.1177/0894318418807931}. The first area we address with this paper is investigating what kind of experiences lead to psychological overwhelm occurring. From this, we form our first research question: \\

\textbf{\rqoneOccur} \\

While our first research question addresses how psychological overwhelm occurs, as in which human experiences lead to it, the second area we address is which experiences humans have when psychological overwhelm does occur. From this, we formulate our second research question:\\

\textbf{\rqtwoExperience}\\

One of our major motivations is investigating the impact of psychological overwhelm on the productivity of software developers. In compliance with the~\gls{ipa} methodology, we assess productivity from personal accounts of our interview subjects. This leads us to our third research question: \\

\textbf{\rqthreeProductivity} \\

Finally, we have come across discussions in which overwhelm is explained as symptoms of stressors being overloaded, leading to an overwhelm of the nervous system, which can cause physical and mental trauma~\citep{StressOverwhelm}. To investigate the impact of stress in relation to overwhelm, we look for emerging themes of stress in our~\gls{ipa} interviews. Our last research question, therefore, is related to stress: \\

\textbf{\rqfourStress} \\

\section{Background and Related Work}\label{sec:RelatedWork}
\glsresetall

A directly comparable work to ours does not exist yet in the field of software engineering, as both the subject is seldom studied~\citep{FeelingOverwhelmedAParsesciencingInquiry} and there are no~\gls{ipa} studies out yet in the discipline.

Furthermore, while performing a literature review of~\gls{ipa} studies in the field of psychology, we found that mentioning related work is not prevalent, as the subject matters are highly specialized. Scholars usually mention frameworks they used briefly and literature regarding the methodology of semi-structured interviews. 

In the following subsections, we present work that we deem as closely related to ours as possible. We begin with work on overwhelm, followed by studies for cognitive overload, as it is closely related to overwhelm. We then present works evaluating productivity of software developers and their emotions, and finally present works using the~\gls{ipa} method.

\subsection{Overwhelm\label{ssec:overwhelm}}

One of the challenges for this paper is forming emerging themes from our interviewees' descriptions of overwhelm. There is no singular, all encompassing definition for overwhelm, as the experience of overwhelm can differ greatly depending on the discipline and people investigated.  

\cite{doi:10.1177/0894318418807931} review scholarly works that investigate overwhelm, from a plethora of different disciplines, with the intent to combine them into emerging themes. Most of these works use qualitative research methods, such as Van Manen's descriptive-interpretive phenomenological method~\citep{gill2020phenomenology} or \citeauthor{parse2016parsesciencing}'s method~\citep{parse2016parsesciencing}. The scholars present three themes to encapsulate all the different findings for overwhelm. Firstly, overwhelm arises as ruin or destruction, which suddenly engulfs a person and gives them a feeling of being smothered, trapped, or drowned. Secondly, overwhelm is accompanied by a feeling of loneliness or isolation, coupled with helplessness and powerlessness. Finally, overwhelm causes to reach for relief, in which a person will attempt to reach out, lash out, or self-harm to cope with their overwhelm.  

\cite{FearOfBeingOverwhelmedAnd} uses a series of descriptors to list feelings associated with overwhelm. These include being fragmented, flooded, smothered, engulfed, drowned, and entrapped.

\subsection{Cognitive overload}

The concept of overwhelm is closely related to the term cognitive overload in workplace context, so work related to this term will be briefly summarized as well.~\cite{kirsh_few_2000} defines cognitive overload in the workplace as a phenomenon caused by the following occurrences:
\begin{itemize}
\item Overly large information supply
\item Overly large information demand
\item Too much need for multitasking and too many interruptions
\item Inadequate workplace infrastructure
\end{itemize}

~\cite{kirsh_few_2000} models supply and demand using information flows that push information onto the individual or pull information from the environment to them.

Pushed information is attributed to methods of spoken or written communication with others or oneself, i.e., memos, letters, newspapers, email, telephone calls, journals, calendars.
Since the paper was published in the year 2000, it does not account for more modern methods of communication, such as instant messengers and social media, which should certainly be also considered.

Unlike pushed information, which the individual has no short-term control over, information is actively pulled by the individual. In practice, this encompasses asking colleagues, collecting information from the World-Wide Web or a library, among countless other examples.

Actively pulling information to the point of facing cognitive overload may sound odd at first.~\cite{kirsh_few_2000} attributes this behavior to an inherent belief in most people, that no matter how much information an individual has collected, additional relevant information is out there, and if collected this instant, saves time in the long term.

\cite{kirsh_few_2000} sees multitasking and interruptions as closely related to workplace infrastructure quality. 

Since multitasking and interruptions are inevitable, a good workplace infrastructure should minimize the time it takes to get to full productivity after an interruption, and reduce interruptions to a level where productivity is not severely affected by them.

\cite{iskander_burnout_2019} looks at cognitive overload from a sightly different angle.
In accordance to what biological research has found out about the human brain, \cite{iskander_burnout_2019} defines the capability of an individual to handle information using two memories. The long-term memory stores information and can handle high volumes of information, while the working or sensory memory, which is used to perform tasks, has comparatively limited capacity.

\cite{iskander_burnout_2019} defines cognitive overload as the phenomenon that the capacity of the working memory is exceeded, and excess information has to be moved somewhere else, e.g., into the long-term memory. As a result of their survey in the field of medicine, they conclude that the symptoms of cognitive overload, namely ``increased
rate of errors, inability to carry out activities to a similar competence as achieved previously'' among others, overlap with symptoms associated with burnout.
Therefore, cognitive overload may be seen as short-term burnout and monitoring it can be used to anticipate burnout.

\subsection{Emotions and software developers' productivity}

Overwhelm is attributed to a variety of negative emotions and feelings, which is why we investigate work related to the impact negative emotions or feelings have on software developer productivity. It has been largely found that positive emotions can contribute to higher productivity~\citep{EmotionsInTheSoftwareDevelopmentProcess, crawford2014influence, graziotin2015How, HappinessAndTheProductivityOfSoftwareEngineers, HappySoftwareDevelopersSolveProblemsBetter}. 

Negative emotions, however, can also pose a risk to productivity. Much research in this field exists, with a variety of different research methods being used.~\cite{anany2019influence} categorize these as based on questionnaires, self-assessment, text analysis, biometric sensors, and emotion inducing. To assess happiness, the Scale of Positive and Negative Experience (SPANE)~\citep{diener2010new} is often used, for example in works by Graziotin et al.~\citep{HappinessAndTheProductivityOfSoftwareEngineers, HappySoftwareDevelopersSolveProblemsBetter, graziotin2018What}. 

\cite{girardi2021emotions} use biometric sensors with popup questionnaires to capture developer emotions.~\cite{EmotionsInTheSoftwareDevelopmentProcess} uses the Job Emotions Scale (JES) as proposed by~\cite{fisher2000mood}.

\cite{oswald2015happiness} attempt to induce positive emotions by showing comedy clips and evaluating with questionnaires. Often, a combination of the aforementioned categories is used to judge emotions and productivity.

\subsection{Interpretive Phenomenological Analysis}

\gls{ipa} is a qualitative method and framework found commonly in psychology and nursing disciplines~\citep{ApracticalguidetousingInterpretativePhenomenologicalAnalysis}. For our study, we looked for related work to answer two questions. First, how~\gls{ipa} has been performed in the past by the scholars that created it, such as~\citep{InterpretativePhenomenologicalAnalysis}. Second, if and how~\gls{ipa} has been used for qualitative studies in the discipline of software engineering.   

\gls{ipa} comes from the field of psychology, which explains why most~\gls{ipa} studies are to be found in that discipline. ~\cite{smith2002risk} investigate the decision-making processes that candidates for Huntington's disease make, or how women's sense of identity shifts when transitioning to motherhood~\citep{smith1999identity}. For these studies, 3 and 5 participants are interviewed, respectively.~\cite{flowers1997health} explore health and romance in the gay community. In the discipline of nursing,~\cite{alexander2004you} investigate women's self-injury in the context of being lesbian or bisexual. While these works give us a good understanding of the workings of an~\gls{ipa} study, the subject matters and disciplines do not come close to our work.

Not much work exists in the field of software engineering using the~\gls{ipa} method, as it is mainly deployed in the areas of psychology and nursing. Computer scientists are, however, occasionally, the subjects of an~\gls{ipa} inquiry, such as in the work of~\cite{huff2017hidden}, in which a student in a computer programming course is evaluated. In this work, there is only one participant.

\section{Methodology}\label{sec:Methodology}
\glsresetall

In this section, we explain our methodology, beginning with a brief over\-view of our chosen research method. We then disclose how we selected our samples for our study. Next, we discuss our data collection and analysis procedures. Finally, we present a section discussing reflexivity concerns with the research team.

\subsection{Research Method}

We are using~\gls{ipa}~\citep{InterpretativePhenomenologicalAnalysis} as the framework to perform a qualitative study in the field of software engineering.~\gls{ipa} aims to investigate how individuals have lived through experiences a researcher wishes to study. To do this, it draws upon principles of phenomenology, which requires research to focus on a person's individual perception and description, and to not immediately try to categorize the person into a predetermined system.~\gls{ipa} requires in-depth examinations of individual cases and necessitates appreciating each person's experiences. Due to this highly focused work, the number of participants is rather low. In~\gls{ipa} research, we do not attempt to test a hypothesis, but rather explore the area of concern to attain an understanding of it~\citep{brocki2006critical}.~\gls{ipa} was introduced to the software engineering field by~\cite{lenberg2023Qualitative}, and we point readers to their study to deepen the overview we provide in the present paper.

\subsection{Sample Selection}

Participants were recruited using an invitation document, which discussed the subject, the methodology, and means of contact. The invitation document was then spread to our work colleagues and acquaintances. The invitation document specifically asked for software developers who had experienced being overwhelmed recently, who work full-time, and whose main occupation is developing software programs. Two participants were found who fit the criteria. Before starting the data collection, we obtained informed consent for participating in the study.

\subsection{Data Collection}
Data were collected using semi-structured interviews~\citep{kallio_systematic_2016}, which is the method most used by~\gls{ipa} studies~\citep{brocki2006critical}. Participants were interviewed in their free-time. The choice of when to place the interview time was given to the participants without restrictions, as we hoped to stifle potential participants hesitations, which could arise from conflicts with their work schedules. Designing the interview was difficult, as many~\gls{ipa} studies do not describe their interview process~\citep{brocki2006critical}. Prompt questions or conscious decisions by the interviewer, stemming from continuous analysis during the interview, are seldom provided. 

In general,~\gls{ipa} studies mostly refer to other theories or writings when describing their interview designs. We, therefore, mostly relied on work related to conducting semi-structured interviews. This led us to create an interview guide with prompt questions, which themselves were sorted into areas of concern. To address issues with related~\gls{ipa} work, we added our interview guide in appendix \ref{appendixA}. 

The interviews were performed online, with participants and the interviewer using a camera and a microphone to communicate.~\cite[OBS]{OBS} was used to record the interviews. The audio track was later isolated and run through the transcription software~\cite{Amberscript}. The transcript was rigorously checked and, when required, corrected.

Each sentence of the transcription was then placed into a cell of a spreadsheet document. The video recording was, subsequently, deleted. Interview transcripts were, as well, deleted after data analysis following local privacy laws.

\subsection{Data Analysis}
Similarly to the issue of data collection, details on the data analysis of related work varies~\citep{brocki2006critical}, with many scholars simply referring to the work by~\cite{smith1999doing}. We base our data analysis methodology on the work by~\cite{wiling2008introducing}. We, however, needed to consider that multiple people were in the field. Through a literature review, we found that, similarly to data collection and analysis,~\gls{ipa} studies performed by a team, in general, do not explore how working in a team influences the~\gls{ipa} method, or mention which role individual team members had in the~\gls{ipa} process. 

Notably, we did find work by~\citet{smith2002risk}, in which they worked in a team and mention the roles of the individual team members. The third author performed the interviews, with the first and the second author performing the initial analysis. In the second analysis stage,~\citet{smith2002risk} exchanged their work, which led to the identification of higher order themes. In the third stage, the second half of the analysis was performed. Additionally,~\citet{smith2002risk} agreed to a set of master themes. 

We modeled our approach to team-based~\gls{ipa} on this framework. Our interview and data analysis team was composed of the second, third, and fourth authors of the present paper. One team member conducted the interviews, with all three performing the analysis. This led to a slight modification of~\citeauthor{smith2002risk}'s team-based approach. The members performed the analysis individually, but in the same spreadsheet document. The labels found in this analysis were immediately clustered, as multiple exchanged and individual works would have been too time intensive.

The analysis itself was divided into four stages. In the first stage, all team members read the first transcript and made brief notes. No labels or themes were assigned, as the first stage is  to get an overview of the transcript. In the second stage, labels are created and assigned to their respective statements. If applicable, statements are assigned multiple labels. After this second stage, the team members exchanged their labels and reviewed them. This leads to the third stage, in which team members collectively clustered themes and named them. In case of contrasts, team members voted on how to proceed. In the final and fourth stage, the results of the synchronized clusters are placed into a summary table.  

These four stages of analysis were repeated for both transcripts obtained from the data collection. In a final stage, the results of both transcript analysis were combined into a singular, resulting table. 

\subsection{Team Reflexivity}

As suggested by~\cite{lenberg2023Qualitative} we use this subsection to further the quality of our qualitative study by providing background information and reflections about the authors. As found by~\cite{tong2007consolidated}, a researcher's personal engagement and possible relation with an interview participant can lead to bias, which must be explored. To allow readers to form their opinion upon the matter, we provide information in alignment with~\citeauthor{tong2007consolidated}'s first domain of reflexivity.

\subsubsection{Personal Characteristics}

\textbf{The second author} was present during the first interview and wrote the protocol. She is doing her master's degree in software engineering at the University of Stuttgart and works part-time in an IT department at an automotive company. The author has successfully completed her bachelor's degree in information systems. She had little to no prior knowledge in the subject.  

\textbf{The third author} was only present during the planning and analysis of the interviews, but not the interviews themselves. He is doing his master's degree in software engineering and has previously completed a software engineering bachelor's degree. Apart from working at the university, developing an application for a research project as the sole developer, the author has neither prior work experience, nor much experience in the subject.

\textbf{The fourth author} conducted the interviews with the participants, and is a master's student of software engineering at the University of Stuttgart. He works as a part-time software developer at a logistics company and has successfully obtained a bachelor's degree in software engineering. He has no prior experience leading interviews, and had, before this work, not much knowledge in the subject.

\subsubsection{Relationships with Participants}

In the case of both participants, a relationship was present before the commencement of the study. The first participant is an acquaintance of the first author, who she met at her former job. The second participant is the cousin of the third author. Both participants were interested in the research topic, but did not indicate any perceivable level of bias or prior assumptions. They were fully aware of the intent and personal goals of the interviewer. Furthermore, the interviewer felt more confident interviewing the second participant due to their close relationship.

\subsubsection{Acquiring Participants}
It was exceedingly difficult to find participants for the study. Over 200 potential candidates saw the interview invitation, with none of them reaching out to participate. When questioning a colleague on why he does not wish to participate, the length of the interview and the research method were cited as the primary concern. We were perceived as mere students, lacking the experience or the authority to conduct interviews on such a personal level and such a delicate subject. The length of the interview would also necessitate sacrificing leisure time, something that most likely deterred participants. This resulted in participants, who are closely acquainted to us, which can be classified as less desirable. Furthermore, while the participants do write code, their main activities at work cannot be classified as writing code. This is also consistent with Hurvich's set of descriptors~\citep{FearOfBeingOverwhelmedAnd}, in which he also includes being smothered or drowned.
\subsubsection{Self-Doubt}
Due to the comments made about us being mere students, the interviewer was sometimes hesitant to challenge, or dig deep, in the interviews. This was more prevalent in the first interview, as the author had no personal relationship with the participant. The interviewer feels that, while valuable insight was gained, a certain barrier was not surmounted. The first participant clearly prepared to give responses in a manner they thought desirable, but even though the interviewer was well aware of this, he was unable to push further into territories the participant perhaps did not plan to discuss.

\section{Results}\label{sec:Results}
\glsresetall

We begin describing our results by presenting the cases individually. For each participant, we have created a table of emerging themes, which have been clustered beneath singular keywords. This was done in accordance with work by~\cite{wiling2008introducing}. 
The themes that are presented are not the work by a singular author, but rather represent a combined view of the individual analysis results of each author. As is typical for~\gls{ipa} studies, direct quotes are included in the presentation~\citep{InterpretativePhenomenologicalAnalysis}. 

After both cases have been presented individually, a third, final subsection will integrate the cases. Each subsection provides the results in relation to our research questions. 

We recruited two participants for our study. James\footnote{We are using fictional names to refer to the participants.} is 44 years old. Charles is 27 years old. Both participants identify as male, they work in IT-based companies, and they develop software as part of their jobs. They do not have management type positions.

\subsection{Single-case analysis: James}

James is a 44-year-old \textit{hybrid software developer}. While he writes code for his job, he also focuses on software testing and the training of others in his field of work. 
He codes at the level of machine code. When asked, he described his workload distribution to consist of roughly 10\% integrating new code, 10\% combining code written by others, 30\% testing code written by others, and the remaining half teaching and forming others. 

He enjoys the variety of his job, but he prefers writing new code. Coding represents to him the  ``getting more experiences'' part of his job. He did not start his work with a university degree, but he attained that later after gathering years of industry experience. 

The results of the analysis of the interview with James can be found in Table \ref{tab:James}. We present all the discovered themes. For each presented theme, we also provide an exemplary interview quote from which the theme originated from.

\begin{longtable}{>{\hspace{0pt}}m{0.221\linewidth}>{\hspace{0pt}}m{0.72\linewidth}}
\caption{Analysis results of James\label{tab:James}}\\ 
\hline
\textbf{Theme}                     & \textbf{Interview Quote}                                                                                              \endfirsthead 
\hline
\textbf{Distractions}              &                                                                                                                       \\
Colleagues                         & “[a distraction] can also be a question from a colleague, but that has to be,   because sometimes I have questions too.”           \\
Customers                          & “A customer in need of consulting is calling, that's really   stressful for me”                                       \\
Digital Communication              & “It can really be my own phone, someone calling or messaging   me.”                                                   \\
Meetings                           & “I see [agile methods] as a double-edged sword, because of too many meetings.”                                     \\
Superiors                          & “You cannot deny a call from a superior”                                                                             \\ 
\hline
\textbf{Types of Overwhelm}        &                                                                                                                       \\
Overwhelm by Variety               & “I would like to focus on one problem”                                                                                \\
Technical Overwhelm                & “The other pillar [of overwhelm] is the technical side”                                                                              \\
Temporal Overwhelm                 & “Sometimes we ran to keep the machines [\dots] alive”                                                                   \\ 
Disturbance Overwhelm              & “A customer calls and that, for me, is stressful”   
\\
\hline
\textbf{Physiological Symptoms}    &                                                                                                                       \\
Concentration Problems             & “I become crippled, and I cannot focus properly anymore”                                                              \\
Feeling of Adrenaline              & “It was a large-scale project, my adrenaline level was really   high”                                                 \\
Hair Loss                          & “I made myself stressed. That manifested in me getting circular   hair loss at the back.”                             \\
Inner Shaking                      & “I'd not say that one starts shaking, but I did feel an inner   shaking”                                              \\
Tiredness                          & “[Whenever I came home] I fell into my bed, having no energy left”                                                    \\ 
\hline
\textbf{Coping}                    &                                                                                                                       \\
Denying Requests                   & “Today I will only focus on three problems, the rest can wait or   give them someone else.”                           \\
Estimating Reasonably              & “That was my main takeaway after 5 years. [\dots] I learned to estimate more reasonably”                                \\
Isolation                          & “I only communicate with one or two people, besides that, I try to   isolate myself.”                                  \\
Reaching for Relief from Superiors & “I need a project manager or someone else, that shields me [from unreasonable requests]”                              \\
Staying Composed                   & “I don't have to put myself under pressure.”                                                                          \\
Work-Life Separation               & “From the first day on, I made it a habit that work has to stay at   work”                                            \\ 
\hline
\textbf{Motivation}                &                                                                                                                       \\
Ambition                           & “The production line has to run, I'm a very ambitious person after   all”                                             \\
Problem-Solving                    & “Problem-solving [is the most fun thing about my work]. [\dots] So not designing new things, but making things run”     \\
Responsibility                     & “the production line, it has to produce products and money”                                                           \\
Sense of Achievement               & “It is very important for me to always have a sense of   achievement.”                                                \\
Technically Exciting               & “Nowadays I only become absorbed by a problem completely, if it is   technically exciting to me”                      \\ 
\hline
\textbf{Pressure}                  &                                                                                                                       \\
Performance Pressure               & “Sometimes there was serious criticism [\dots] there are superiors that are absolutely evil to build up pressure”       \\
Self-Induced Pressure              & “I put myself under pressure, It has to work out, no matter   how”                                                    \\
Social Pressure                    & “You want to make a good impression at first, and put yourself under   pressure”                                      \\
Subconscious Pressure              & “Outlook\footnote{Microsoft's E-Mail client.}, for example, always pops up notifications for messages [\dots]. That creates subconscious pressure.”          \\ 
\hline
\textbf{Emotions}                  &                                                                                                                       \\
Anger                              & “It pisses me off, because I have a clear goal, but my environment   does not notice that”                            \\
Hesitation                               & “I would not call it fear, but I did have hesitations [to tell my chef about failures]”                               \\
Frustration                        & “It's a combination. I'm sad but also frustrated. [\dots] Sadness does predominate though.”                             \\
Sadness                            & “I'm not sure if sadness is the right word, I want to cry, it   goes like shit again”                                 \\ 
\hline
\textbf{Origins of Overwhelm}      &                                                                                                                       \\
Criticism from Superiors           & “There also is overwhelm by variety, but that comes with the job   criticism from superiors I have”                   \\
Failing Colleagues                 & “One colleague does not manage to complete a task, how do I explain   that to the customer in a diplomatic way”       \\
Heavy Workload                     & “We worked ten hours a day, the maximum amount possible [by law].”                                                             \\
Old Code                           & “One example [for a time-consuming task] is me getting very old code, which is the case right now”                    \\
Position                           & “There also is overwhelm by variety, but that comes with the job   position I have”                                   \\
Reassignment                       & “You also have to complete Task B\footnote{The participant referred to a \textit{Task B} to mean unexpected, unplanned, and undesired tasks.} [\dots], the superiors put a gap into what you're doing”                              \\
Variety of Tasks                   & “There were too many small problems, and the machine had to   run”                                                    \\ 
\hline
\textbf{Personality}               &                                                                                                                       \\
Aversion to Management             & “I once was team leader [\dots] where I had to mediate between [stakeholders]. [\dots] I decided to never become a boss”  \\
Maturity                           & “I have become more mature and more thick-skinned”                                                                    \\
Others Are Just Like Me            & “at one point you realize, the others also put their pants on   one leg at a time”                               \\
Youthful Naivety                   & “Back then when you're young and also in a new company and   environment, you are more careful”                       \\ 
\hline
\textbf{Productivity Impactors}    &                                                                                                                       \\
Being in The Zone                  & “If you want to develop software [\dots], you have to immerse yourself in your own nerd world”                          \\
Decreased Productivity             & “A lot of time is lost [during interactions with a superior], you could design the process differently”               \\
Mental Focus                       & “When you build a framework [for problem-solving] in your head [\dots] and it is torn down by a call, that costs time”  \\
Silence                            & “I am the type of guy, that sometimes puts ear protection on, if I   want to immerse myself”                          \\
Workflow                           & “There is the interruption that really drags you out of your workflow, [\dots] you have to get back into the problem”   \\ 
\hline
\textbf{Demographic}               &                                                                                                                       \\
Education                          & “I completed a German Realschulabschluss\footnote{Part of the so called \textit{Mittlere Reife} (Middle Maturity) of the German schooling system, it is oriented to occupation rather than an academic career.}. [\dots] [later] I studied electric engineering”                                 \\ 
\hline
\textbf{Impact}                    &                                                                                                                       \\
Communication                      & “There was an impact on communication. regarding the way I   accepted additional work”                            \\
Reflection                         & “For clinical reasons I said 'Stop, you really need to step down a   gear, and take a breather.'”                    \\
Sleep                              & “If I get an idea, then I get out of bed, at night. That did happen   to me already, I write down a note”             \\
Stress (Negative)                             & “I believe it's definitely a sign of stress, If you carry your work   around with you 24/7”                           \\
Taking Work Home                   & “Wherever you think I'd be relaxed [\dots] my processor\footnote{As in, my own CPU, meaning ``my brain''.} was always in the highest gear, trying to find a new solution”      \\
\hline
\end{longtable}

We will now present the results of James' interview in relation to the research questions.

\subsubsection{\rqoneOccurShort} 

James had prepared himself for the interview beforehand, as when asked about his experiences with overwhelm, he stated that ``overwhelm [\dots] can be described by two pillars''. The first pillar he mentioned was \textit{Temporal Overwhelm}, the second being \textit{Technical Overwhelm}.

\textit{Temporal Overwhelm} was a recurring theme in our discussion, as it was often stated that the limited amount of time he had to spend on ``too many small problems'' led to James feeling overwhelmed. When asked about technical difficulty regarding feeling overwhelmed, James stated that ``there is no problem that is too difficult, the problem is how much time do I have for the problem'', which indicates that technical difficulty---which James defined as \textit{Technical Overwhelm}---leads back to \textit{Temporal Overwhelm}. We prompted to elaborate further on \textit{Technical Overwhelm}. One type of technical task that overwhelms is working on old code, which makes things even worse when the authors of said code no longer work for the company. \textit{Temporal Overwhelm} thus stems from two fundamental issues, being the amount of tasks a developer has, and the difficulty of the tasks in relation to how much time they require to complete. \textit{Temporal Overwhelm} also describes periods in time, in which James had to work ``ten hours a day'', which is the upper legal limit to work in a day in Germany, for long periods of time, up to five years. The sheer amount and length of work time feels overwhelming. 

While these are the types of overwhelm James had thought of before the interview, more origins for feeling overwhelmed could be found when interviewing him. One of the major emerging themes of the interview was the role distractions play in feeling overwhelmed. James says that ``if you want to develop software [\dots] you have to immerse yourself in your nerd world'', meaning that, to be able to maintain mental focus at work is paramount for his mental well-being. Distraction, be it from ``customers in need of consulting'', or ``calls from superiors you can't deny'', causes loss of mental focus, which  when occurring repeatedly, leads to feeling overwhelmed by others. 

Overwhelm by demanding people was also reported. James, however, only experienced this type of overwhelm when he slipped into the role of management once, and suddenly ``was [\dots] the firewall between the customer and the developers''. Recalling his experiences, he states ``everybody was getting in my face, which was also why I decided I never want to be the boss''. Feeling overwhelmed can also stem from being reassigned to new work, as ``you have pressure, you need to do task B [\dots] because your boss hits you with a real gap in your work''. This ``really annoyed'' James. This leads back to the loss of mental focus or ``getting in the zone''. James describes reassignment through variety, stating ``variety can be overwhelming; however, that's the nature of my position''. He would rather ``focus on one task''. 

\subsubsection{\rqtwoExperienceShort} 

James maintained that the experience of overwhelm ``is very dependent on the personality'', and differs between younger and older, more experienced developers. Furthermore, the reaction to the aforementioned different sources of overwhelm are divergent.  

Starting with \textit{Temporal Overwhelm}, James found that in the early days, he was ``young and naïve'', which resulted in him having as mentality that ``I have to accomplish this'' coupled with large amount of adrenaline. He ``put [himself] under pressure'', thinking ``this has to work out, no matter what''. This further perpetuated a state of constant overwhelm, especially when working on large projects, resulting in ``adrenaline levels that were extremely high''. James noted, however, that physiological symptoms started to show, as he states: ``I made myself stressed. That manifested in me getting circular hair loss''. He also observed colleagues battling with overwhelm, as ``in the project with 50 programmers, [they had] 4 burnouts, people that were gone for good''. He attributes himself not getting burned out with his youthful naivety. He saved himself from further physiological harm by performing a reflection of his situation, realizing ``I need to go down a gear, take a breather''. 

His experience of \textit{Temporal Overwhelm} has changed over the years, which he attributes to himself having matured as a developer. He has adopted multiple coping strategies, which include denying requests, telling others ``today I will only focus on three problems, the rest can wait''. He attributed his experiences gained in estimating his work tasks as a major relief, as his ``main takeaway after 5 years'' is that he ``learned to estimate more reasonably.''. Emotional maturity also plays a role in preventing \textit{Temporal Overwhelm}, as he taught himself to ``not put [himself] under pressure''. Generally, when younger, he felt a sense of responsibility to ``keep the systems running'', a feeling that gradually decreases with maturity. 

As a younger developer, \textit{Temporal Overwhelm} can cause a sense of dread, as one is hesitant to report failures or give notice on incomplete tasks to superiors. James notes that relief from superiors is vital for preventing pressure building up on developers, as superiors ``are vital'' and ``shield [\dots] from unreasonable requests''. During \textit{Temporal Overwhelm}, reaching for relief from superiors seems to be an important part in keeping developers from reaching their breaking points. While, as a mature developer, James rarely feels the pressures he faced as a young developer, he now mostly reacts with a mixture of frustration and sadness, as \textit{Temporal Overwhelm} makes him think ``I could cry right now, [\dots] not this shit again''. 

Overwhelm caused by distractions or reassignments, which we call \textit{Disturbance Overwhelm}, is experienced differently, as the main emotional response by James is a self-described mixture of anger and frustration. Reassignment of task and role create a ``zone of emotional tension''. This is further nurtured by an aversion to management intervention, which take James out of the zone and his mental focus. James, again, deals with this through his more mature interactions with his surroundings, as he is now capable to tell his boss ``get off my back'' and others ``sorry, this counter is now closed''. This is something he was incapable of doing as a younger developer, as he ``was more careful'' of his surroundings and ``these days I can just reflect stuff back''. To his colleagues, he is more lenient, as he sometimes needs help from them. To this, he says: ``It can also be a question from a colleague [which is distraction], but that has to be, because, sometimes, I have questions too''. With growing maturity, he stopped trying to please everybody, as he realized that ``the others also put their pants on one leg at a time''. More coping mechanism used to prevent or lessen the effects of \textit{Disturbance Overwhelm} is self-isolation, as James would purposefully ``only communicate with one or two people''. He is also easily disturbed by his cellphone or notifications by his company's internal messaging system. He states that the constant pop-ups create subconscious ``pressure''. To ensure minimal disturbances, James would also use hearing protection as not to hear any noises around him and thus isolate more effectively.

In general, overwhelm caused James to take his work home, and even in times when he tried to relax his ``processor was always in the highest gear, trying to find a new solution''. This could cause James to lose sleep, as he would ``get out of bed, at night'' and ``write down a note'' to solve a problem. James stated he tried to form ``a habit that work has to stay at work'' from ``the first day on'', but it was clear that this was not always successful.  

\subsubsection{\rqthreeProductivityShort} 

James found that his productivity was, sometimes, not impacted by \textit{Temporal Overwhelm}, as it filled him with adrenaline. This helped him ``get into the zone'', an important factor for his productivity. Short periods of \textit{Temporal Overwhelm} led James to perceived increased productivity. It is unclear how this would have been in the long term, though, as James self-reflected and acknowledged that he needed to ``slow down'' to prevent any more physiological symptoms.  

\textit{Disturbance Overwhelm}, on the other hand, was blamed for a massive loss in productivity. Any form of distraction would disturb James mental focus, and repeated disturbance would lead to James feeling ``pissed off''. James says distractions ``pull [him] out of the workflow'', requiring him to ``think himself back into the problem''. Communication with colleagues could also suffer, as James would sometimes reply to his colleagues in ways that ``could be seen as unfriendly''. This was magnified when James thought the colleague calling him was incapable.

\subsubsection{\rqfourStressShort} 

Stress was mentioned by James throughout the interview, without direct prompt by the interviewer. Stress played a pivotal role in James self-appointed pressure during \textit{Temporal Overwhelm}. While putting himself under pressure due to his ambition, he
``induced the stress'' to himself, a byproduct of trying to cope with his state of overwhelm. Stress also plays a part in \textit{Disturbance Overwhelm}, as James repeatedly mentioned feeling stressed, stating ``a customer calls and that, for me, is stressful''. Furthermore, James states that ``internal messengers, such as Teams and Skype, really cause stress''. When this type of stress accumulates, James reports a ``crippling'' feeling, as if ``shaking internally''. It breaks his concentration, therefore impacting the quality of his work. 

When recalling his time in management, he remembers ``psychic stress'', a unique experience which stemmed from having to tolerate obvious ``negative statements'' from others. In his current position, he, as previously mentioned, can deny requests or reflect back to the person making the statement, something he could not while in a management position. James also described his mind constantly trying to ``review situations'' and ``find new solutions'' outside work hours as a sign of stress.

\subsection{Single-case analysis: Charles}

The results of the analysis of the interview with Charles can be found in Table \ref{tab:Charles}. As with James, we present the discovered themes accompanied by an exemplary interview quote.

Charles is a 27-year-old IT-Consultant who primarily works in advising banks on technical issues. Being an IT-Consultant, Charles is mainly deployed as an analyst. His business is understanding technical problems that banks have, collecting data, and then using low-code~\citep{sahay2020supporting} development platforms to develop solutions. Charles, hence, develops software at the opposite pole of abstraction level than James, who writes code that is nearer to the binary code of CPUs. While writing  classical code very little, Charles writes software to test and validate his solutions. Charles holds a Master's degree in  business information technologies, and he started working directly after completing his degree. He worked for the same company part-time while he was a student. 
 
\begin{longtable}{>{\hspace{0pt}}m{0.226\linewidth}>{\hspace{0pt}}m{0.714\linewidth}}
\caption{Analysis results of Charles\label{tab:Charles}}\\ 
\hline
\textbf{Theme}                      & \textbf{Interview Quote}                                                                                                   \endfirsthead 
\hline
\textbf{Distractions}               &                                                                                                                            \\
Colleagues                          & “Most of the time, someone wants something from me multiple times a day.”                                                  \\ 
\hline
\textbf{Types of Overwhelm}         &                                                                                                                            \\
Communication Overwhelm             & “[\dots] I was overwhelmed, the first time I was in that situation [having to call other departments or even customers].”                                                         \\
Disturbance Overwhelm               & “[when colleagues, customers, or managers] want something from me multiple times a day for something else[\dots] this is definitely something, I would say, that contributes to overwhelm.”                                           \\
Organizational Overwhelm            & “I believe it is rather that [organizing], which creates overwhelm and not the actual tasks [\dots]”                         \\
Positive Overwhelm                  & “I believe a healthy measure of overwhelm is quite pleasant [\dots] pressure, to a degree, is very positive for me.”         \\
Technical Overwhelm                 & “[\dots] you feel overwhelmed, thinking, I am too stupid for this.”                                                          \\
Temporal Overwhelm                  & “At some point it just becomes exhausting because you are working a lot.”                                                 \\ 
\hline
\textbf{Physiological Symptoms}     &                                                                                                                            \\
Sleep Loss                          & “My manager told me once that he could not sleep one night, as he was so stressed out.”                                    \\ 
\hline
\textbf{Coping}                     &                                                                                                                            \\
Accepting Limited Knowledge         & “We work on a level of complexity where you quickly learn, it is not worth it to dig into some things.”                    \\
Asking for Help                     & “One realizes quickly that it is ten times more efficient to just accept that you need to ask someone for help [\dots]”      \\
Isolation                           & “I usually start at eight, it helps, most people start at nine. That is one hour of peace and quiet.”                      \\
Learning to Prioritize              & “If something is more important than something else, that other thing is just dropped.”                                    \\
Learning to Say No                  & “[\dots] I can not help you right not, like I generally can help you, but not right now.”                                    \\
Not My Problem                      & “At the end of the day, if it all goes to shit, it is not my problem.”                                                     \\ 
\hline
\textbf{Mitigating factors}                 &                                                                                                                            \\
Experienced Colleagues              & “I have no idea what I am supposed to do, then someone else looks at it and tells me he has already seen that ten times.”  \\
Isolation from Management Decisions & “Things like budgetary concerns never reach us.”                                                                           \\
Non-Competitive Work Culture        & “If things were more me-first orientated, I believe it would be very uncomfortable [working].”                             \\
Others Are Just Like Me             & “[\dots] you feel stupid until you realize, everybody is feeling like me.”                                                   \\
Relief from Superiors               & “[\dots] our team leader is very good in actively keeping external pressure from reaching the team [\dots].”                   \\ 
\hline
\textbf{Motivation}                 &                                                                                                                            \\
Ambition                            & “I somewhat have ambition to ensure we accomplish stuff and I stand for the things that come.”                             \\
Problem-Solving                     & “We do also realize very complex things, and those are the things that are fun for me to do.”                              \\
Responsibility                      & “[Being unproductive] It costs money, and not a small amount of money”                                                  \\ 
\hline
\textbf{Emotions}                   &                                                                                                                            \\
Annoyance                        & “[Me and others] are not exactly overwhelmed [\dots] but are annoyed to death.”                                                \\
Confusion                           & “At some point, sometimes, you think to yourself, I do not understand why this is not working.”                            \\
Feeling Competent                   & “[Being asked for help] is also something positive [\dots] it is a sign, that you are doing a good job [\dots].”               \\
Frustration                         & “Most of the time, everybody around you is just as pissed off, annoyed, and frustrated as yourself.”                       \\ 
Self-doubt                         & “Sometimes when you start stuff, you feel overwhelmed, because you are thinking to yourself, you are just too stupid.”    \\
\hline
\textbf{Origins of Overwhelm}       &                                                                                                                            \\
Bad Code                            & “We are working only with bad work.”                                                                                           \\
Bad Documentation                   & “[\dots] the whole time we are practically only working on stuff that is documented badly or incorrectly.”                   \\
Bad Tasks                           & “This [the task] is just crap.”                                                                                            \\
Being New                           & “When you come into a new project [\dots] at the beginning you are always [\dots] overwhelmed [\dots].”                          \\
Crunch                              & “[\dots] we were somewhat, I believe, on 9 to 10 hours.”                                                                     \\
Deadlines                           & “If you have a go-live [\dots] in July [\dots] it was clear in May that it will be very stressful.”                            \\
Having No More Ideas                & “There are moments, in which you are overwhelmed because you just run out of ideas.”                                      \\
Pressure                            & “There have been moments where we thought, okay, this is going to get close.”                                              \\
Reassignment                        & “It can happen, that tomorrow you are told, forget what you have been working on the last two weeks.”                      \\ 
\hline
\textbf{Personality}                &                                                                                                                            \\
Youthful Naivety                    & “One tends to [when young] initially not talk about things that are bad [\dots] one wants to be solution oriented [\dots]”     \\ 
\hline
\textbf{Productivity Impactors}     &                                                                                                                            \\
Information Overload                & “Starting off there is just an enormous amount of information that you need to know, in order to work productively.”       \\
Organizing Priorities               & “[\dots] prioritize yourself and when in doubt, if you are overwhelmed with prioritizing, simply delegate it.”               \\
Organizing To-Dos                   & “I have somewhat roughly organized To-dos, but that is already super helpful [\dots] to keep an overview.”                   \\
Unavailable Colleagues              & “I cannot do anything against it, you stand there doubting yourself, until people have time that can help you.”           \\ 
\hline
\textbf{Impact}                     &                                                                                                                            \\
Personal Limits                     & “At some point one tells themselves, okay, i have enough of this!.”                                                        \\
Stress (Negative)                 & “You have people who give themselves stress, and I definitely put myself under more stress.”                              \\

Stress (Positive)                      & “[I find it] somewhat pleasant, as it helps me to focus better.”                                                         \\
Taking Work Home                    & “[\dots] I don't necessarily take the stress home with me, but rather the thinking about it.”                                \\
Work-Life Separation                & “At some point we started discussing, if it would be required that people work on the weekends.”                          \\
\hline
\end{longtable}

\subsubsection{\rqoneOccurShort} 

From the interview with Charles, we could extract multiple types of overwhelm. Starting with \textit{Temporal Overwhelm}, Charles described times in which he simply became overwhelmed by exhaustion from ``working a lot''. He simply had ``too many small tasks'' and was ``overwhelmed from the sheer mass''. The tasks themselves also contributed to feeling overwhelmed, as Charles is only ``working with bad work'' (meaning, bad architectural choices or low-quality code). This includes tasks Charles considers as ``just crap'', and tasks which are ``documented badly or incorrectly''. 
Time pressure is exerted when deadlines have to be met, which leads to developers having to work long hours, with Charles stating he worked ``9 to 10 hours'' a day. \textit{Temporal Overwhelm} was also magnified when getting close to the aforementioned deadlines, by superiors asking if it is ``necessary that people work on the weekends'', giving Charles the feeling of ``okay, this is looking really bad''. 

While having too many tasks leads to \textit{Temporal Overwhelm}, Charles also described it leading to \textit{Organizational Overwhelm}. When having too many tasks, Charles felt that it was not ``the challenge to complete them all'', but rather to ``get it all organized''. The organization of the tasks was what ``really created the overwhelm'', as one first ``needs a system'' to process the workload. Even with the tasks themselves, Charles finds it sometimes difficult to know where to start, a further symptom of bad documentation.  

When requiring help, Charles sometimes had to call other departments or even customers, an experience in which he was ``overwhelmed the first time I was in that situation''. Especially with customers, he is cautious not to ``do or say anything negative''. Charles recalled times in which he ``said something he was not supposed to say'' or ``awkwardly worded'' something. While he can handle these situations, he mentions that some colleagues have ``major stress'', indicating different personalities might experience this type of overwhelm more severely.  

Another emerging theme of overwhelm is \textit{Disturbance overwhelm}. Charles likes digging into singular, large problems, but is frustrated when people ``want something from me multiple times a day for something else''. Generally, consistent calls from colleagues are overwhelming. Charles recalled when his company's division in Austria did not share a holiday with the German side, and the Austrians reported the next day they ``have never been so productive in months''. This led to internal discussions in management to perhaps limit disturbances among colleagues. However, Charles also stresses the importance of being able to get help from experienced colleagues, calling it ``super helpful'', when a technical expert was solely placed into their project to answer questions.

The final emerging theme was that of \textit{Positive Overwhelm}. This was mostly tied to \textit{Temporal Overwhelm}, with Charles stating that ``a healthy measure of overwhelm is quite pleasant''. He finds that while being overwhelmed with things to do, it helps him ``to focus''.

\subsubsection{\rqtwoExperienceShort} 

Charles experienced the different kinds of emerging overwhelms in different ways. Charles described a \textit{Positive Overwhelm}, a state in which he placed himself under immense amounts of pressure. This pressure was not perceived as negative though, as it helped him focus more on his work, with him even describing the feeling of being ``pleasant''. 

Very different was his experience of \textit{Technical Overwhelm} and \textit{Disturbance Overwhelm}, in both cases associating negative emotions to the state of being. Charles often described tasks which were, in his opinion, of poor quality, and working on such tasks could trigger a wide range of emotions, including annoyance, with him and his colleagues being ``annoyed to death'' by strenuous tasks. Bad tasks could also lead the developers to feel anger or, as Charles described it, being ``pissed off''. Bad code or bad documentation induced \textit{Technical Overwhelm} also lead to moments of helplessness and confusion, as Charles state one simply ``runs out of ideas''. Struggling with problems made him ``feel [\dots] overwhelmed because you are thinking, you are just too stupid''. \textit{Technical Overwhelm} also stemmed from Charles simply being new to his job or project, as one ``in the beginning [\dots] is always overwhelmed [\dots]''. 

The experience of \textit{Communication Overwhelm} was described as different based on the underlying personality, but Charles mostly felt anxiety that he may accidentally ``say the wrong thing''. 

The experiences tied to \textit{Organizational Overwhelm} can mostly be tied to the symptoms following bad organization, which is confusion, with Charles thinking ``what am I supposed to do now''. Singular, large problems seem to be preferable, and not being able to start due to bad organization puts Charles in a sort of limbo where he can never be fully productive. Trying to keep an overview of one's tasks was described to cause one to ``lose energy'' quickly. For these situations, it was crucial for Charles to dispense of his overwhelm through various means.  

The first method with which he could cope with growing overwhelm was learning to ask for help. Especially at the beginning, Charles described himself as trying not to cause any problems. ``One tends not to talk about things, when they are shit [\dots] because you are trying to be solution oriented and positive and bla bla bla [\dots]''. This tendency to please did Charles no good, as he quickly ``came to the realization, that it is ten times more efficient, to just accept that you should ask somebody, instead of trying to dig into the problem yourself''. Trying to dig into a problem himself led Charles to quickly ``not know where to start''. For him, it was a key point in his, so far, short career, to realize that he needs to ``give up the idea to understand everything'' and rely on ``people externally''. Part of this realization is accepting that, when somebody you need is not available, then ``that is dumb, but you cannot do anything about it [so accept it]''. This ties into Charles accepting limited knowledge, the fact that he cannot and will not be an expert of everything. He works on ``a level of complexity, where you quickly learn that it is not worth digging into stuff yourself''. It was part of a learning process, as he initially had the ambition to ``do it'' himself and ``wanted to understand it'' himself. 

In relation to \textit{Organizational Overwhelm}, Charles has learned to deal with it through improved prioritization. Learning to decide ``what is more important right now'' and, when in doubt, asking others to prioritize for oneself, has been important. Learning to say no to his colleagues has helped with \textit{Disturbance Overwhelm}, even though Charles still occasionally gets annoyed, especially when he thinks someone ``should know that himself''. He also starts his workdays earlier in an attempt to get ``an hour of peace and quiet''. Finally, Charles has distanced himself from the responsibilities that he is not directly responsible for, such as the financial situation of his company or deadlines promised to customers from management, simply stating that ``[..] if it goes to shit, it's really not my problem''. He is not worried about problems that might ``cause some unknown person to lose his bonus''.

During the interview, Charles mentioned several factors that mitigated symptoms of the occurrence of overwhelm. The most important of these was the relief provided by management. He attributes his team lead's ability to ``actively keep external pressure away from the team'' as one of the causes of more productive work, as many problems simply ``do not reach you''. Monetary concerns or thoughts of ``are we taking too long'' did not reach him. Superiors would also prioritize for him when he did not know how to himself. He described it as a ``feeling of freedom'' when receiving direction from his superior. Furthermore, his superior would ``clearly tell'' him what to do and shield him from requests by others, telling them that ``no, you have to wait''. In general, Charles feels that his superiors are very concerned and ambitious to keep pressure from reaching the developers. He does note that they are ``paid very well for this''. 

Another important mitigating factor was Charles's easy access to experienced colleagues, as not having this would have increased his self-pressure and helplessness when tackling problems he could not understand. Charles also feels that his colleagues and him form a good team. He attributes a lack of me-first attitudes as making his work environment more pleasant. Generally, his colleagues mostly originate from STEM fields, which he attributes to people being less competitive as, for example, ``law or business students''. Charles often struggled with self-doubt regarding his abilities, questioning himself if he is ``too stupid''. This was alleviated when he looked around himself and realized, ``everybody around me is suffering just like me, it is not just me''. This clicked for him when he once worked with his manager, who ``for 20 minutes, did nothing but curse'', helping him understand that he is ``not the problem''.

Charles did not suffer any physiological symptoms from feeling overwhelmed; however, he reported that his manager had lost sleep over the stress he felt. He did state that he was pushed to his personal limits, which led to him telling himself ``I have had enough of this''. When overwhelmed, Charles did take work home, meaning his thoughts were still at work while at home.

\subsubsection{\rqthreeProductivityShort} 

Information overload from starting on a new project is overwhelming for Charles, as there is ``a lot of information you need to know, to work productively''. Receiving too much information at once or too many tasks thus leads to reduced productivity. As previously mentioned, too much time is lost when developers have to deal with external stressors or pressure from management decisions which do not impact them directly, that is why productivity was lost when that did happen to be the case. Disturbances are a major source of productivity loss, as the most productive Charles and his colleagues have ever been was when one department which usually disturbs them was on vacation. Charles noted that when digging into a complicated problem, he needs around 20 minutes to be mentally focused enough to start being productive. Any disturbance during or after this time requires another 20 minutes, thus hindering him from being productive. When under \textit{Positive Overwhelm}, Charles felt that the stress he was receiving was pleasant, he was more productive than without it. This may be due to him being able to achieve mental focus more easily or disregard negative symptoms resulting from his personal ambitions and self pressure.

\subsubsection{\rqfourStressShort} 

Stress was mentioned by Charles without prompt by the interviewer, indicating a correlation between feeling stressed and overwhelmed. In summary, two types of stress were described, positive stress and negative stress. The symptoms of negative stress were not elaborated upon further by the interviewee, but generally, in all situations in which Charles felt overwhelmed, he also felt stressed. Charles described the stress he inflicted upon himself as positive, as it helped him be more productive.

\subsection{Integration of cases}

To compare the experiences of both participants with feeling overwhelmed, we enlist the themes that emerged from their respective interviews in Table~\ref{tab:Summary}. If an interviewee has provided information belonging to a certain theme, the theme is marked with a $"+"$ in the table. Conversely, themes that are not touched upon in a particular interview are marked with a $"-"$. 

On closer examination, some parallels in the experiences of both participants can be observed. There is, however, a non-trivial number of differences in their recollections. The parallels and the differences are summarized later on in the section, in correspondence to the research questions of this work.

\begin{longtable}{>{\hspace{0pt}}m{0.643\linewidth}>{\centering\hspace{0pt}}m{0.138\linewidth}>{\centering\arraybackslash\hspace{0pt}}m{0.138\linewidth}} 
\cline{1-2}
\textbf{Distractions}               &     & \multicolumn{1}{>{\hspace{0pt}}m{0.138\linewidth}}{}  \endfirsthead
Colleagues                          & $+$ & $+$                                                   \\
Customers                           & $+$ & $-$                                                   \\
Digital Communication               & $+$ & $-$                                                   \\
Meetings                            & $+$ & $-$                                                   \\
Superiors                           & $+$ & $-$                                                   \\ 
\hline
\textbf{Types of Overwhelm}         &     & \multicolumn{1}{>{\hspace{0pt}}m{0.138\linewidth}}{}  \\
Communication Overwhelm             & $-$ & $+$                                                   \\
Disturbance Overwhelm               & $+$ & $+$                                                   \\
Organizational Overwhelm            & $-$ & $+$                                                   \\
Overwhelm by Variety                & $+$ & $-$                                                   \\
Positive Overwhelm                  & $-$ & $+$                                                   \\
Technical Overwhelm                 & $+$ & $+$                                                   \\
Temporal Overwhelm                  & $+$ & $+$                                                   \\ 
\hline
\textbf{Physiological Symptoms}     &     & \multicolumn{1}{>{\hspace{0pt}}m{0.138\linewidth}}{}  \\
Concentration Problems              & $+$ & $-$                                                   \\
Feeling of Adrenaline               & $+$ & $-$                                                   \\
Hair Loss                           & $+$ & $-$                                                   \\
Inner Shaking                       & $+$ & $-$                                                   \\
Sleep Loss                          & $-$ & $+$                                                   \\
Tiredness                           & $+$ & $-$                                                   \\ 
\hline
\textbf{Coping}                     &     & \multicolumn{1}{>{\hspace{0pt}}m{0.138\linewidth}}{}  \\
Accepting Limited Knowledge         & $-$ & $+$                                                   \\
Asking for Help                     & $-$ & $+$                                                   \\
Denying Requests                    & $+$ & $-$                                                   \\
Estimating Reasonably               & $+$ & $-$                                                   \\
Isolation                           & $+$ & $+$                                                   \\
Learning to Prioritize              & $-$ & $+$                                                   \\
Learning to Say No                  & $-$ & $+$                                                   \\
Not My Problem                      & $-$ & $+$                                                   \\
Reaching for Relief from Superiors  & $+$ & $-$                                                   \\
Staying Composed                    & $+$ & $-$                                                   \\
Work-Life Separation                & $+$ & $-$                                                   \\ 
\hline
\textbf{Mitigating Factors}                 &     & \multicolumn{1}{>{\hspace{0pt}}m{0.138\linewidth}}{}  \\
Experienced Colleagues              & $-$ & $+$                                                   \\
Isolation from Management Decisions & $-$ & $+$                                                   \\
Non-Competitive Work Culture        & $-$ & $+$                                                   \\
Others Are Just Like Me             & $-$ & $+$                                                   \\
Relief from Superiors               & $-$ & $+$                                                   \\ 
\hline
\textbf{Motivation}                 &     & \multicolumn{1}{>{\hspace{0pt}}m{0.138\linewidth}}{}  \\
Ambition                            & $+$ & $+$                                                   \\
Problem-Solving                     & $+$ & $+$                                                   \\
Responsibility                      & $+$ & $+$                                                   \\
Sense of Achievement                & $+$ & $-$                                                   \\
Technically Exciting                & $+$ & $-$                                                   \\ 
\hline
\textbf{Pressure}                   &     & \multicolumn{1}{>{\hspace{0pt}}m{0.138\linewidth}}{}  \\
Performance Pressure                & $+$ & $-$                                                   \\
Self-Induced Pressure               & $+$ & $-$                                                   \\
Social Pressure                     & $+$ & $-$                                                   \\
Subconscious Pressure               & $+$ & $-$                                                   \\ 
\hline
\textbf{Emotions}                   &     & \multicolumn{1}{>{\hspace{0pt}}m{0.138\linewidth}}{}                                                     \\
Anger                               & $+$ & $-$                                                   \\
Annoyance                           & $-$ & $+$                                                   \\
Confusion                           & $-$ & $+$                                                   \\
Hesitation                          & $+$ & $-$                                                   \\
Feeling Competent                   & $-$ & $+$                                                   \\
Frustration                         & $+$ & $+$                                                   \\
Sadness                             & $+$ & $-$                                                   \\ 
Self-doubt                         & $-$ & $+$
\\
\hline
\textbf{Origins of Overwhelm}       &     & \multicolumn{1}{>{\hspace{0pt}}m{0.138\linewidth}}{}  \\
Bad Code                            & $-$ & $+$                                                   \\
Bad Documentation                   & $-$ & $+$                                                   \\
Bad Tasks                           & $-$ & $+$                                                   \\
Being New                           & $-$ & $+$                                                   \\
Criticism from Superiors            & $+$ & $-$                                                   \\
Crunch                              & $-$ & $+$                                                   \\
Deadlines                           & $-$ & $+$                                                   \\
Failing Colleagues                  & $+$ & $-$                                                   \\
Having No More Ideas                & $-$ & $+$                                                   \\
Heavy Workload                      & $+$ & $-$                                                   \\
Old Code                            & $+$ & $-$                                                   \\
Position                            & $+$ & $-$                                                   \\
Pressure                            & $-$ & $+$                                                   \\
Reassignment                        & $+$ & $+$                                                   \\
Variety of Tasks                    & $+$ & $-$                                                   \\ 
\hline
\textbf{Personality}                &     & \multicolumn{1}{>{\hspace{0pt}}m{0.138\linewidth}}{}  \\
Aversion to Management              & $+$ & $-$                                                   \\
Maturity                            & $+$ & $-$                                                   \\
Others Are Just Like Me             & $+$ & $-$                                                   \\
Youthful Naivety                    & $+$ & $+$                                                   \\ 
\hline
\textbf{Productivity Impactors}     &     & \multicolumn{1}{>{\hspace{0pt}}m{0.138\linewidth}}{}  \\
Being in The Zone                   & $+$ & $-$                                                   \\
Decreased Productivity              & $+$ & $-$                                                   \\
Information Overload                & $-$ & $+$                                                   \\
Mental Focus                        & $+$ & $-$                                                   \\
Organizing Priorities               & $-$ & $+$                                                   \\
Organizing To-Dos                   & $-$ & $+$                                                   \\
Silence                             & $+$ & $-$                                                   \\
Unavailable Colleagues              & $-$ & $+$                                                   \\
Workflow                            & $+$ & $-$                                                   \\ 
\hline
\textbf{Demographic}                &     & \multicolumn{1}{>{\hspace{0pt}}m{0.138\linewidth}}{}  \\
Education                           & $+$ & $-$                                                   \\ 
\hline
\textbf{Impact}                     &     & \multicolumn{1}{>{\hspace{0pt}}m{0.138\linewidth}}{}  \\
Communication                       & $+$ & $-$                                                   \\
Stress (Negative)                    & $+$ & $+$                                                   \\
Stress (Positive)                    & $-$ & $+$                                                   \\
Personal Limits                     & $-$ & $+$                                                   \\

Reflection                          & $+$ & $-$                                                   \\
Sleep                               & $+$ & $-$                                                   \\
Taking Work Home                    & $+$ & $+$                                                   \\
Work-Life Separation                & $-$ & $+$                                                   \\
\hline
\caption{Analysis results of JAMES and Charles\label{tab:Summary}}
\end{longtable}

\subsubsection{\rqoneOccurShort} 

James and Charles list a variety of overwhelm origins. The quality of the code the participants are working with can be recognized as a common origin of overwhelm for both. Furthermore, both participants mention being reassigned from one task to another as an additional reason. While using different terminology, parallels between both participants could be recognized in terms of crunch and heavy workload. Both report the feeling of overwhelm being amplified by distractions such as colleagues.

\subsubsection{\rqtwoExperienceShort} 

Both participants have negative experiences with \textit{Temporal Overwhelm} and \textit{Technical Overwhelm}. These are the only two types of overwhelm that both interviews have in common. 
While other types of overwhelm were also mentioned by both participants, only Charles had some experiences with overwhelm with positive outcomes. 

Due to feeling overwhelmed, both interview participants report experiencing a range of mostly negative emotions. The only emotion which was commonly reported by both subjects was the feeling of frustration. While the emotional consequences of overwhelm were shared by both participants, only James reported having physiological symptoms like hair loss, excessive tiredness and concentration problems. Feeling overwhelmed impacted the work-life balance of both subjects, who both report taking work home at a certain point. Other than that, the feeling of overwhelm caused both participants to be stressed, with only Charles sometimes seeing this in a positive light. 

To fight the feeling of overwhelm, both participants have developed several coping mechanisms. Denying requests and isolating themselves from distractions are the only common ones. When it comes to outside help, both James and Charles view their superiors as stress mitigating factors. While Charles reports being able to passively rely on his superior, James had to develop the coping mechanism of actively seeking help from his superiors.

\subsubsection{\rqthreeProductivityShort} 

To both participants, being overwhelmed can have positive and negative impact on their productivity. Overwhelm can boost their ability to concentrate and get in the zone, to be able to fight \textit{Temporal Overwhelm}. 

\textit{Technical overwhelm} on the other hand, manifested as a multitude of different tasks for example, leads to a decrease in productivity. 

\subsubsection{\rqfourStressShort} 

Both interviewees talked about stress without being prompted by the interviewer. This points toward the supposition that stress is closely related to the feeling of being overwhelmed. 
A reoccurring type of stress that both participants connect to being overwhelmed is the self-inflicted stress. While in the case of James, it could be viewed negatively, Charles reported it having a positive effect on his productivity.

\section{Discussion}\label{sec:Discussion}
\glsresetall

In this section we discuss our findings regarding the overwhelm, productivity, and stress. Each area of interest is discussed in a separate section, beginning with overwhelm.

\subsection{Overwhelm}
We set out to explore the experiences of developers feeling overwhelmed, a goal which we have achieved through our analysis. We have discovered multiple emerging themes of overwhelm, namely communication, disturbance, organizational, variety, technical, temporal, and positive overwhelm. 

After researching overwhelm in the scientific literature, we were keen to see how overwhelm in the discipline of software engineering overlaps with other disciplines such as psychology. \citet{doi:10.1177/0894318418807931} performed a literature review and formalized three themes of overwhelm, to which we will compare our findings.

The first theme by~\citet{doi:10.1177/0894318418807931} is that of overwhelm arising as a form of ruin or destruction, which suddenly engulfs a person and gives them a feeling of being smothered, trapped, or drowned. This theme also partially emerged from our interviews, with James describing the stress caused by overwhelm to have a ``crippling'' feeling, similar to the descriptors of drowning or smothering. Charles did not use such descriptors to explain his experience of overwhelm. In general, they did not describe overwhelm as suddenly engulfing them, but rather overwhelm stemming from constant and consistent pressure, resulting in stress and, ultimately, overwhelm. Our participants mentioned that they have witnessed colleagues who experienced total ruin in the form of burnouts and sleep loss. We are not collecting the theme of burnout in our results, as our participants did not report having experienced it.

The second theme is that of overwhelm being accompanied with loneliness or isolation, with a feeling of helplessness or powerlessness. This theme did not emerge as obviously from our interviews as the previous one, however, parts of it could be found. Charles described a feeling of helplessness, being lost, not knowing what to do, when being overwhelmed. He did not, however, feel lonely or isolate himself as a result. The same can be said for James, from whom none of the themes mentioned above emerged. James was, however, proud of his position as a veteran developer, and perhaps did not want to disclose any moments of weakness to the much younger and inexperienced interviewer. Another likely scenario is that, due to his experience, James simply has not felt overwhelmed in a way described above in a while, and simply can not recall his experiences from way back when. It is, of course, possible that he indeed never had experienced overwhelm with feelings of loneliness or isolation. James and Charles both had colleagues and superiors who helped them cope with their overwhelm, so perhaps the symptoms stated above only appear when there is nobody who can help you. The second theme of overwhelm not occurring prominently may be due to unique properties of the discipline of computer science, with employees being less social, therefore reducing me-first mentality. This may, however, lead to deficits in other categories.

The third theme is that of overwhelm causing subjects to reach for relief, doing so by either reaching out to others, lashing out at others, or self-harming. These mechanisms help to cope with their overwhelm. We also found parts of these themes emerging from our interviews. 
For both participants, dealing with bad tasks or unclear prioritization of tasks could be alleviated by reaching for relief from superiors. Charles believes that his job would be much less enjoyable without the ability to reach out for help. This also includes help from colleagues, which is also the case for James. While both acknowledge that colleagues disturbing them is annoying, at the same time, they both are tolerant towards it, as they themselves reach out to colleagues for relief. James and Charles both mentioned that they started saying no to their sources of disturbances to prevent themselves from becoming overwhelmed. 
James, in particular, mentioned that his communication with his colleagues became a little more hostile, however, one can not really call this lashing out. None of the participants mentioned lashing out at people in their private life due to overwhelm.

\subsection{Productivity}
Before conducting the interviews, we expected the participants to report productivity loss with overwhelm; however, we did not expect overwhelm being associated with gains in productivity. While James did not explicitly state that overwhelm had a positive effect on productivity, he described putting himself under pressure and focusing more when overwhelmed. In this state, he felt a strong ambition to continue working and solve the problem, evoking a sort of responsibility for it. Being able to be ambitious about his problem and gaining mental focus for a single problem was described by James to be beneficial for his productivity. Overwhelm, therefore, could have contributed positively to productivity here. Charles explicitly stated that overwhelm caused by numerous tasks, strict deadlines, or a difficult problem, is pleasant, as it increased his self-pressure and made him work more effectively. This discovery would not coincide with the findings by~\citet{akula2008impact}, who found that stress leads to bad code and less productivity. We cannot, however, compare our results quantitatively. 

What James and Charles describe is similar to crunch time~\citep{edholm2017crunch}, though not as intense, as neither of the participants worked unpaid overtime or weekends. According to~\citet{edholm2017crunch}, this time of extreme working can be seen as good or bad, with people seeing it as a sometimes necessary evil. It may very well be that the participants both view themselves as being more productive, but in reality are working sloppy and are unproductive. James and Charles both mention, however, that there is a limit to positive overwhelm, as at some point both needed to stop themselves to prevent any physiological or psychological symptoms. Both participants stated the importance of being able to reach out to either colleagues or superiors, with Charles admitting that if those means of relief were not present, he would be impacted more, which could lead to less productivity. 

Our most fortified finding regarding productivity impact is from the negative implications of disturbance overwhelm. Both James and Charles reported that their productivity was impacted massively by distractions from colleagues, superiors, or in James case, digital communication. They describe that there is a certain level of mental focus one must achieve to work productively, with Charles specifically stating around 20 minutes for a complex problem. Any repeated disturbances take the developers out of their focus, requiring a period of time for them to regain it. While not the worst offender, reassignment also seems to ruin productivity, as the developer is still working on the previous task in his mind, and needs time to readjust. As Charles mentioned, it is difficult to be told to drop a problem one has been working on for two weeks and be expected to perform well on the new task immediately. Only one of them mentioned that too frequent meetings are also a distraction.

\subsection{Stress} 

Psychological stress can be described as emotional pressure or strain~\citep{moore1996stress}, which might explain why the participants mentioned stress so often without being prompted by the interviewer. Generally, when looking at the interview transcript, many occurrences of stress could be exchanged with pressure, and many occurrences of pressure with stress. When asked about stress, the participants struggled to describe what it is for them, but mainly the theme of pressure emerged. Of all negative states that were mentioned, stress was the most frequent. This may be because the participants simply struggled to describe which emotions they experienced during overwhelm, and diverted to stress, which encompasses many potential emotional strains.

\section{Conclusion}\label{sec:Conclusion}
\glsresetall
In this paper, we report on an~\gls{ipa} study on experiencing overwhelm in a software development context. Throughout a qualitative analysis of the shared experiences, we uncover  seven categories of overwhelm:
\begin{itemize}
    \item \textit{Communication Overwhelm} describes developers being overwhelmed when confronted with critical social situations. 
    \item \textit{Disturbance Overwhelm} describes developers being overwhelmed by disturbances, be it from meetings, colleagues, reassignment, or noise. 
    \item \textit{Organizational Overwhelm} describes when developers are overwhelmed by the task of prioritizing and organizing their workloads. 
    \item \textit{Variety Overwhelm} describes developers being overwhelmed by the sheer number of tasks. 
    \item \textit{Technical Overwhelm} describes developers being overwhelmed by the difficulty or size of singular tasks. 
    \item \textit{Temporal Overwhelm} encompasses all feelings of being overwhelmed from running out of time or not being able to allocate time. 
    \item \textit{Positive Overwhelm} describes developers feeling ambitious from being overwhelmed, making them perceive the overwhelm as a positive thing.
\end{itemize}

From all categories, \textit{Disturbance Overwhelm}, \textit{Temporal Overwhelm}, and \textit{Positive Overwhelm} seem to impact the developers' productivity the most, with occurrences of disturbance and temporal overwhelm leading to reduced productivity. Positive overwhelm was described by the participants to nurture productivity by self-pressure and ambition.

Participants would cope with overwhelm by accepting that they do not need to know everything, and simply accept when they should drop a task or ask others for help. If requests overwhelmed the participants, they have learned to deny requests they know would overwhelm them, something that only came with maturity, since they, as younger developers, tried to solve every problem and not complain. Better estimations and learning to organize priorities also prevented overwhelm. One of the most important coping mechanisms was reaching for relief from colleagues or superiors who would, in the case of technical overwhelm, help them with their expertise or, in the case of organizational overwhelm, prioritize tasks for them.

Factors that mitigated or prevented overwhelm were also discovered, such as the important role of management in shielding the participants from external pressure. This can include budget concerns or deadline issues. Realizing that the other people are just like oneself, and that some tasks are perceived by everybody as bad, no matter the experience, helped to mitigate the participants feeling stupid or inept. Interestingly, participants mentioned the non-competitive nature of their work environment, which may be a unique factor mitigating overwhelm themes, such as loneliness, that commonly occur in disciplines such as nursing. Finally, a lack of feeling any responsibility or consequence was also reported as mitigating overwhelm. If the company goes bankrupt, it may be unpleasant, but at the end of the day, the participants are not personally impacted besides having to look for a new job. Perhaps developers, who have more personal stake in the company they are working for, are overwhelmed more frequently.

Stress was the universal descriptor when talking about the experience of overwhelm, however, frustration, sadness, anger, annoyance, confusion, fear, and anxiety were also reported. Stress was used to generalize emotional strain, which the participants experienced when overwhelmed.

\subsection{Outlook}

This work took merely a small glimpse into the experiences developers have when overwhelmed. While we have discovered interesting results, our interviews also uncovered areas of interest for further research. Our participants showed many of the same emerging themes of overwhelm as participants from other disciplines; however, we found that the unique properties our participants described in their workplaces, being a non-competitive culture and a disconnect from external pressures, may have mitigated many of the overwhelm experiences. 

As it is not part of~\gls{ipa} studies to generalize results, we call future studies in software engineering research to design qualitative and quantitative studies that would explore further our findings.

Future work could investigate the experiences of overwhelm of team leaders or senior developers, perhaps of some who have personal stakes in the company. Our participants were never worried about budget or deadline concerns, as they have no personal stake in the projects. Our participants, furthermore, mentioned that they have experienced colleagues or superiors, who had strong physiological and psychological reactions to being overwhelmed, such as strong fatigue, the inability to sleep, and burnout. Researchers could attempt to acquire participants who have suffered from these symptoms. It seems personality plays an important role in how overwhelm is experienced and perhaps some can handle it much worse than others, for example, our participants seem to handle overwhelm quite well.

\appendix

\section{Interview Guide}
\label{appendixA}
 \vspace{1cm}
\begin{enumerate}
\setcounter{enumi}{0}
    \item[\textbf{0.}] \textbf{ Ease Interviewee into the interview}
    \begin{enumerate}
        \item[--] Interviewer reveals general information about him/herself
            \begin{enumerate}
                \item[$\circ$] Student of Software Engineering at University of Stuttgart
                \item[$\circ$] Doing this as part of a research project alongside regular lectures
                \item[$\circ$] Inform participant of data collection, risks and benefits, and rights
                \item[$\circ$] Small talk with interviewee\\
            \end{enumerate}
    \end{enumerate}
    \item[\textbf{1.}] \textbf{Explaining what the study wants to investigate (Overwhelm, especially during programming)}
    \begin{enumerate}
        \item[--] Experiences during overwhelm (emotions, thoughts, etc.)
         \item[--] Origin and effect of overwhelm\\
    \end{enumerate}
    \item[\textbf{2.}] \textbf{Explaining our methodology}
    \begin{enumerate}
        \item[--] We conduct interviews with multiple employees
         \item[--] The content is audio recorded via cellphone and will be transcribed with a software 
         \item[--] Afterwards, 3 students read the transcriptions and summarize parts of it using keywords
         \item[--] Then, the summaries of all students are compared and the keywords are refined to match between them\\
    \end{enumerate}
    \item[\textbf{3.}] \textbf{Asking for Background Information}
    \begin{enumerate}
        \item[--] Gender
         \item[--] Age
         \item[--] Education
         \item[--] Job Title
         \item[--] Role in the Development Process
         \item[--] Which part of the software do you usually work on?
         \item[--] How much of your work is actually programming, as opposed to management, coordination, etc.
         \item[--] What are your other job tasks apart from programming?
         \item[--] Which one of your job tasks do you prefer more and why?
         \item[--] How long have you been working as a developer?\\
    \end{enumerate}
    \item[\textbf{4.}] \textbf{Overwhelm in general}
    \begin{enumerate}
        \item[--] Overwhelm is not uncommon and is present in any field of work
         \item[--] A broad definition: ”Too stressed out to complete work, even though willing to do it” 
         \item[--] Example scenario where people usually feel overwhelmed (e.g., ”Imagine you are expected to change a part of a large software, which you have never looked at in your life”, ”Onboarding: You enter a new company and are expected to deliver prompt results”)\\
    \end{enumerate}
    \item[\textbf{5.}] \textbf{Lived experience of overwhelm}
    \begin{enumerate}
        \item[--] What experiences have you made of being overwhelmed?
         \item[--] Ask interviewee to remember situations where he felt overwhelmed
         \item[--] Ask interviewee to describe what overwhelm feels like \\
    \end{enumerate}
    \item[\textbf{6.}] \textbf{Origin and effect of overwhelm}
    \begin{enumerate}
        \item[--] How often do you feel overwhelmed?
         \item[--] When do you usually feel overwhelmed? (In what kind of situations do you feel overwhelmed? What are the reasons for that?)
         \item[--] What causes could your overwhelm have? Is it from the code specifically or other circumstances?
         \item[--] Can a single cause be identified or is it a composition of different causes?
         \item[--] What effect does feeling overwhelmed have on your communication with people around you? (co-employees, family, friends)
         \item[--] What effect does this feeling have outside of your working hours/in your free time?
         \item[--] Does overwhelm impact your productive capabilities? If  so, how?
         \item[--] How do you deal with being overwhelmed?
         \item[--] Have you spoken with colleagues of yours about being overwhelmed?
    \end{enumerate}
\end{enumerate}

\bibliographystyle{elsarticle-harv} 
\bibliography{michels2024-overwhemled_devs_ipa-techreport}

\end{document}